# Thresholdless Coherent Light Scattering from Subband-polaritons in a Strongly-Coupled Microcavity.

Johannes Gambari[1], Antonio I Fernandez-Dominguez[1], Stefan A Maier[1], Ben S Williams[2,3], Sushil Kumar[3], John L Reno[4], Qing Hu[3] and Chris C Phillips[1]

[1] *Physics Dept., Imperial College London, London, SW7 2AZ UK.*

[2] *Dept of Electrical Engineering and California NanoSystems Institute, University of California, Los Angeles, California 90095, USA.*

[3] *Massachusetts Institute of Technology, Dept. of Electrical Engineering and Computer Science and Research Laboratory of Electronics, Cambridge, Massachusetts, 02139, USA.*

[4]*Sandia National Laboratories, Department 1123, MS 0601, Albuquerque, New Mexico, 87185-0601, USA.5*

We study a “strongly-coupled” (SC) polariton system formed between the atom-like intersubband transitions in a semiconductor nanostructure and the THz optical modes that are localised at the edges of a gold aperture. The polaritons can be excited optically, by incoherent excitation with bandgap radiation, and we find that they also coherently scatter the same input laser, to give strikingly sharp “sideband” (SB) spectral peaks, in the backscattered spectrum. The SB intensity is a sensitive track of the polariton density and they can be detected down to a quantum noise floor that is more than 2500 times lower than the excitation thresholds of comparable quantum cascade laser diodes. Compared with other coherent scattering mechanisms, higher order SB scattering events are readily observable, and we speculate that the effect may find utility as a passive all-optical wavelength shifting mechanism in telecommunications systems.

In a strongly-coupled (SC) system, the electron and photon sub-systems are so strongly coupled to each other that they form a new hybrid entity, a polariton, where the coupling is characterised by a vacuum-Rabi energy, $\hbar\Omega_{VR}$, that exceeds the original natural linewidths of both the electron and photon modes. Viewed in the time domain, the excitation energy cycles coherently, at a rate $\hbar\Omega_{VR}^{-1}$ between electronic and photonic forms, before being lost, either by optical emission or by other non-radiative loss channels [1].

Here we study a SC system whose polaritons are formed from the THz photon modes of a tightly-confined metal-semiconductor microcavity, that are hybridised with the electronic “Intersubband transitions” (ISBT’s) in a semiconductor nanostructure. These polaritons carry an optical dipole, so when we create a population of them incoherently, (with an interband near-infrared pump laser, $\omega_{NIR}$ (fig.1)), we find that they coherently scatter that same input beam, generating strikingly sharp, multiple-order sidebands (SB’s), at $\omega_{SB} = \omega_{NIR} + n\omega_{THz}$ (n=-2,-1,0,+1+2) in the spectrum of backscattered light.

The devices studied were fabricated from epilayers comprising many (~175) repeats [2,3] of a GaAs/ [Al,GaAs] semiconductor nanostructure module [fig 1]. The epilayers were fabricated into ~10μm thick gold-epilayer-gold sandwich waveguides that used the near ideal gold conductivity [4,5] to confine the THz fields into layers ~ 10 times thinner than the λ ~ 100 μm free-space wavelength. Three device structures were studied, identified here as “λ~80μm” [3], “λ~100 μm”[2] and “λ~120” μm[3] corresponding to the approximate THz free-space wavelengths they emitted when they were electrically driven, in separate experiments [2,3], as standard quantum cascade lasers (QCL’s).

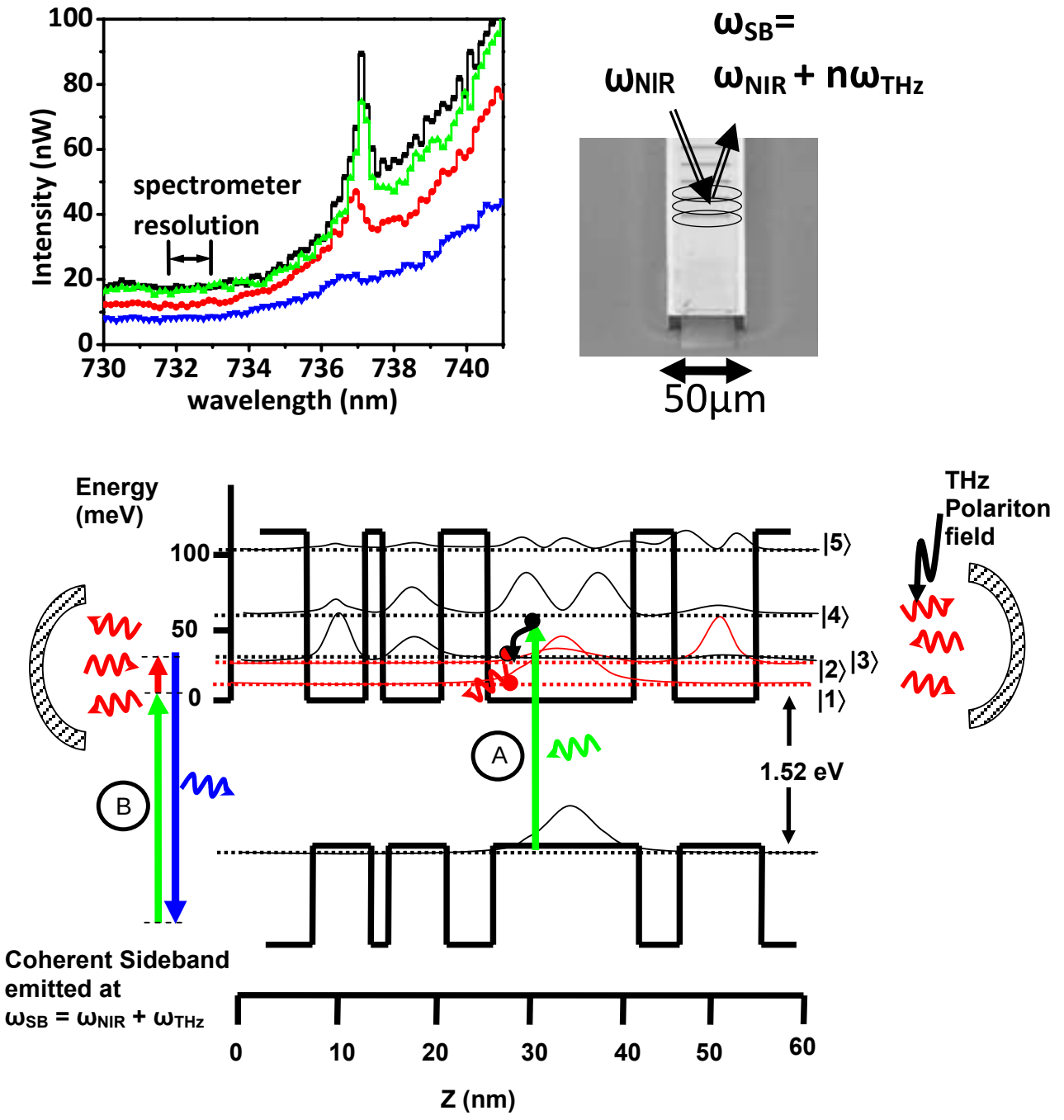

**Fig 1.** A single period of the nanostructure, embedded in a THz microcavity which is symbolised here by external mirrors. THz polaritons are created (A) by photo carriers, excited by a near IR laser, $\omega_{NIR}$, cascading down through the subband system until they reach the polariton state. The same near IR laser (B), is also coherently scattered from these $\omega_{THz}$ polaritons, and generates “sidebands” at $\omega_{SB} = \omega_{NIR} + n\omega_{THz}$. Upper left inset:- Raw spectra of the light backscattered from the top of the device, as the ~50 μm diameter λ~744nm $\omega_{NIR}$ laser spot is scanned along the centre of the ridge, across one of the slots, at distances from the centre of the slot of 0 μm (squares) , 20μm (triangles), 40μm (circles) and 100μm (inverted triangles).

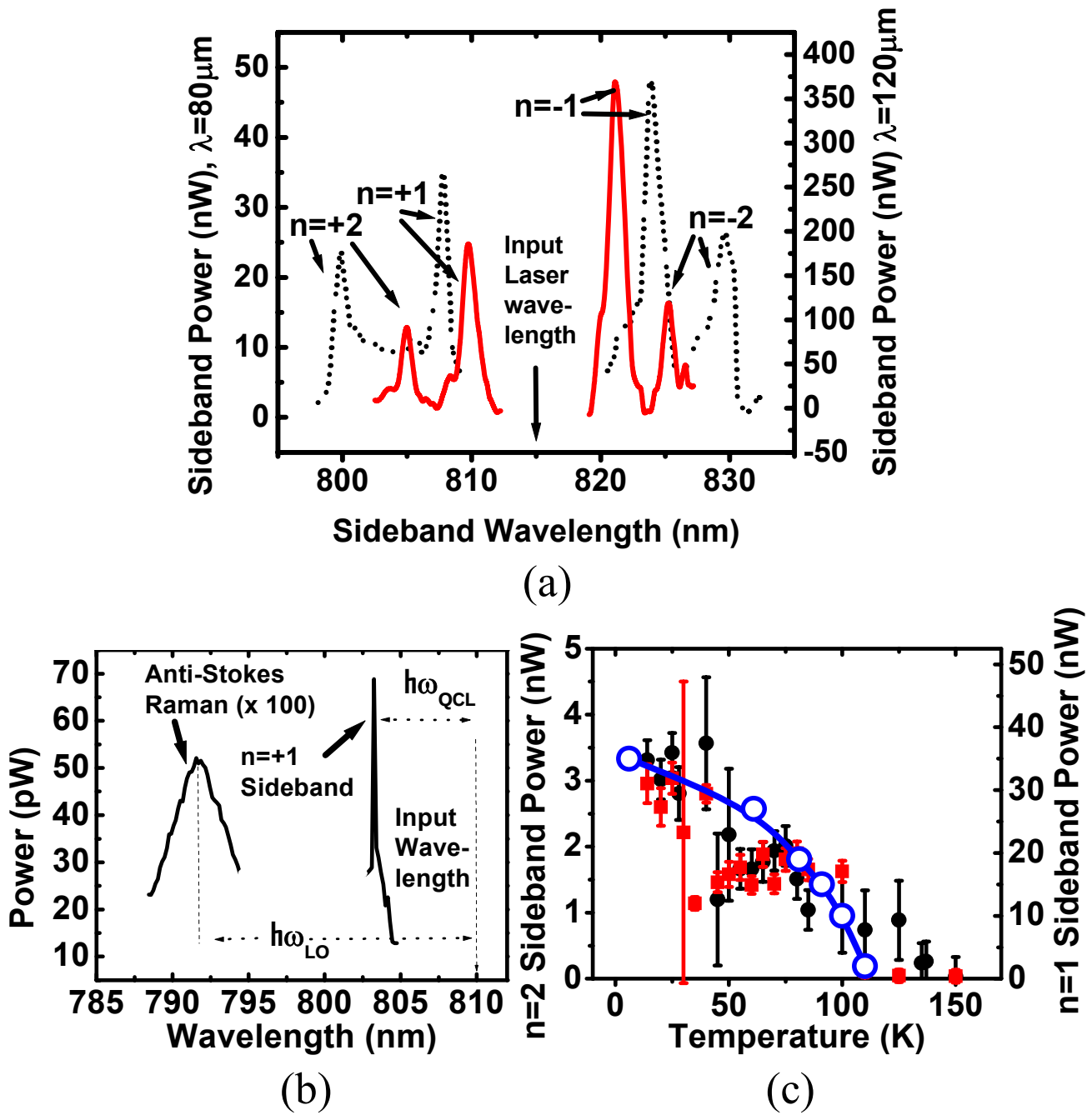

Fig 2.(a) T~14K Background-subtracted spectra taken, from the "λ~120 μm" (solid line, right hand vertical axis), and "λ~80 μm" (dotted line, left hand vertical axis) devices. Both spectra show coherent sideband features at $\omega_{SB} = \omega_{NIR} + n\omega_{THz}$. (n=-2,-1,+1,+2), whose linewidths are set by the spectrometer resolution [Δλ~ 1.4nm in this case]. (b) Comparison between the T~ 14K Anti-Stokes GaAs Raman line, [enhanced by x100 on this plot] and a typical n=+1 sideband feature. (c) Temperature dependence of the SB intensity measured with a near IR wavelength of 810nm and ~1mW power entering the "λ~80 μm" device; squares, the n=-1 SB intensity ( right hand axis), filled circles, the n=-2 sideband intensity (left hand axis); open circles, output of a QCL made from the same heterostructure [3] (arb units).

Each period of the the "λ~100μm" nanostructure in fig. 1 [2] comprises **4.9**/7.9/**2.5**/6.6/**4.1**/15.6/**3.3**/9.0 nm thick layers of $Al_{0.3}Ga_{0.7}As$/GaAs and supports ~5 confined electron states. The 15.6 nm well is doped at 1.9 x $10^{16}$ $cm^{-3}$, giving an areal electron density, $n_s$ = 3 x $10^{10}$ $cm^{-2}$ and a Fermi energy of ~1meV, so only the lowest subband is occupied at equilibrium. The 1.52 eV photoluminescence peak implies an effective bandgap close to that of bulk GaAs and the |2> ⇒ |1> ISBT had a modelled energy of $E_{12}$ =15.8 meV and a transition dipole of $z_{12}$ = 2.3 nm. The waveguide is ~10μm x 50 μm x 834 μm long. It has a periodic array (fig. 1 inset) of 6 30 μm x 8 μm slots, spaced by Λ = 31 μm, etched into it's top surface. These were originally designed to outcouple the THz fields when the structure was used as a QCL [4].

The devices were illuminated normally, with a tuneable continuous-wave titanium:sapphire near-infrared laser, $\omega_{NIR}$, with a 50μm spot size and a linewidth $\Delta\omega_{NIR}$ < 0.1 meV. The backscattered light was polarisation filtered before being analysed with a 0.25m grating monochromator and a standard, background-subtracting, photon-counting setup whose cooled photomultiplier had a dark count < 10 cps. The sharpness of the SB's meant careful attention to the system's mechanical and laser wavelength stability was needed to see them.

Backscattered spectra were typically dominated by an elastically scattered background peak and a λ =815 nm / 1.52 eV photoluminescence peak (not shown), but they also featured strikingly sharp "sideband" (SB) peaks. The data presented in figures 1 - 3 were taken with the spectrometer slits opened up to improve the signal-to-noise ratio, but separate trials (not shown) always found SB linewidths that were resolution-limited, down to the 0.3nm/0.5meV working limit of the spectrometer. For a given device, the SB features stayed at a fixed frequency interval from $\omega_{NIR}$, as the Ti:sapph laser was tuned over a wide wavelength range (fig.3).

All the SB's disappeared if $\hbar\omega_{NIR}$ was tuned below the ~1.52 eV/ 815 nm effective bandgap of the nanostructure. Both 1st and 2nd order peaks vanished [fig.2 (c)] by ~120K, similar to the maximum operation temperature of comparable QCL's [3]. The optical polarisation of the SB's matched the polarisation of the $\omega_{NIR}$ input to better than 99%, whether it was linear or circular.

When a small bias was applied across the waveguide it acted as a photoconductive detector which could be calibrated to measure the fraction of the $\omega_{NIR}$ beam that made it through the illuminated slot in the top gold layer. When the laser spot was scanned along the centre of the waveguide, across a given slot, (Fig 1 inset) the SB intensity scaled with this photocurrent.

The SB's were reproducible over months of measurement and through numerous changes to the optical setup. They were never seen in control experiments, when the $\omega_{NIR}$ spot was focussed onto (i) un-metallised epilayer samples, (ii) onto GaAs test wafers, (iii) onto highly scattering parts of the cryostat mount or (iv) onto the gold contacting layers adjacent to the slots in the structure of fig.1.

The way the SB's tune with $\omega_{NIR}$ [figs. 3(b) and 3(c)] immediately implies that they are generated by a form of coherent scattering. We rule out normal phonon Raman scattering because the SB's are ~ 100 x more intense, at least 12 times sharper, and appear at the wrong energy offsets [fig. 2(b)] compared with typical LO phonon Raman lines.

Coherent scattering processes generate SB's whose lineshape is a convolution of the linewidths of the input laser (<0.1meV) and that of the scattering excitation. Therefore, the sharpness of the SB lines [< 0.5meV] compared with an estimated bare ISBT linewidth of ~ meV [6] argues that they cannot be due to standard electronic Raman scattering from the ISBT.

SB features which are superficially similar to the ones we see here have previously been generated by exploiting optical frequency mixing effects which arise due to non-linear components of the optical polarisability in the material of a QCL [7,8]. However, this mechanism requires the presence of a spectrally sharp THz optical field inside the structure, and this could only be happening here if we had somehow managed to produce an optically-pumped

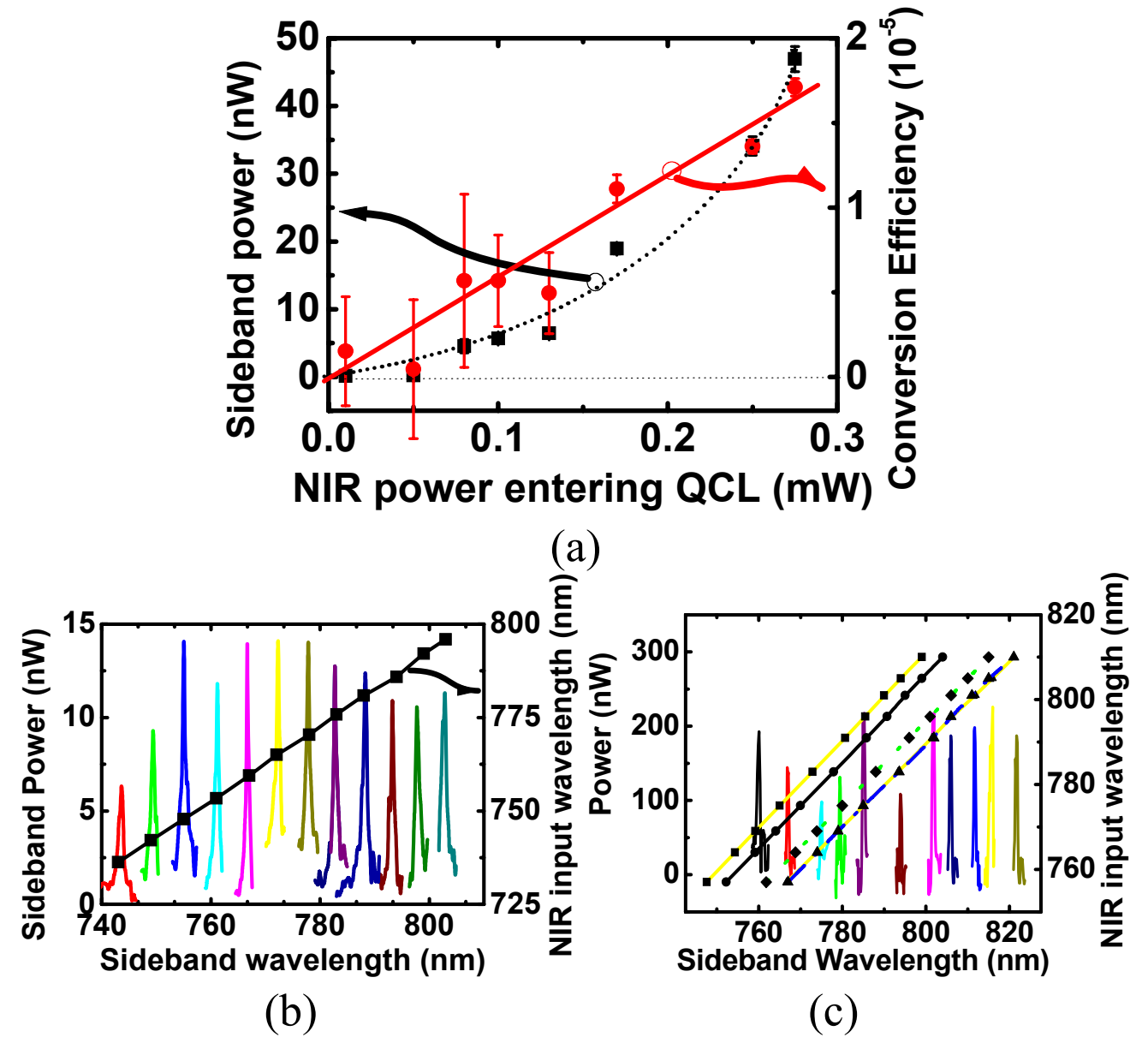

Fig.3 (a) Sideband power, (squares, left hand axis) and conversion efficiency (circles, right hand axis) from the "λ~100μm" device of fig.1. The Dashed (solid) lines are guides to the eye, evidencing the quadratic(linear) dependencies of the power (efficiency) at low NIR pump levels (b) tuning behaviour (right hand axis) and n=1 sideband spectra (left hand axis) of the "λ~100 μm" structure as the near IR input wavelength (1mW power) is scanned . (c) tuning behavior (right hand axis) of the n= +2 (squares), n=+1 (circles), n=-1 (diamonds) and n= -2 (triangles) SB peaks from the "λ~80 μm" structure. Spectra (left hand axis), of the n=-2 SB's peaks from the same sample at the input wavelength denoted by the corresponding triangle data point on the tuning line. All data taken at T~14K.

analogue of a standard QCL. This possibility, however, is strongly at odds with the fact that we see no excitation threshold for the onset of the SB generation (Fig.3). Defining the SB generation efficiency, $\eta_{SB}$ , as the ratio between the detected sideband intensity and the $\omega_{NIR}$ power entering the slot [Fig.3(a)], $\eta_{SB}$ climbs linearly from the detector system's photon noise floor as the $\omega_{NIR}$ pump power is increased. When a comparable structure to the device of fig. 1 was configured as a QCL and electrically excited [2], it's threshold current density, $J_{th}$ ~ 435 A $cm^{-2}$, corresponded to ~4.8 x $10^{27}$ $sec^{-1}$ $m^{-2}$ ISBT transitions throughout the epilayer's 175 periods. In our optical experiments (fig.3) the SB first appears with ~0.1mW entering the 2.4 x $10^{-10}$ $m^2$ slot, an areal excitation rate ~1.7 x $10^{24}$ $sec^{-1}$ $m^{-2}$ , i.e. ~2500 times smaller than the comparable QCL threshold.

Finally, the ease with which the n=±2 higher order processes are seen [fig.2(a)] contrasts sharply with what is observed in Raman and non-linear frequency mixing experiments.

To understand the origin of the sidebands we first need to estimate the coupling between the ISBTs and the photon modes in these devices. The ISBT energy is independent of electron in-plane wave vector, so strong electron correlation effects [9] concentrate all the oscillator strength into a single, dispersionless, atom-like Lorentzian line, with a large transition dipole, $z_{12}$ , that has already been shown to generate SC with giant $\hbar\Omega_{VR}$ energies in planar structures [10,11], especially with the wider wells used in THz devices [12]. In fact, $\hbar\Omega_{VR}$ values have been achieved that not only exceed the linewidths, but are also are comparable with the transition energy itself [13,14], the so-called ultra-SC (USC) condition[15].

In our non-planar devices, the photon modes are confined in all 3 dimensions, so to estimate the degree to which they couple with the ISBT's we must first compute their mode shapes and volumes. This is done with a standard finite-difference-time-domain (FDTD) calculation, on a 125 nm mesh, which treats the gold as a perfect conductor and the semiconductor as an insulator with a dielectric constant of 13.3 [16]. It models the ridge structure of fig 1 as an infinite array of slots so that periodic boundary conditions allow the modes to be plotted in terms of the superlattice wavevectors, $2\pi/\Lambda$, where $\Lambda$ = 31 μm is the slot repeat distance (fig.4)

The model correctly reproduces the "radiating" modes (fig. 4(a) squares) that were originally intended to outcouple [4] the THz when biased as a QCL. However, at

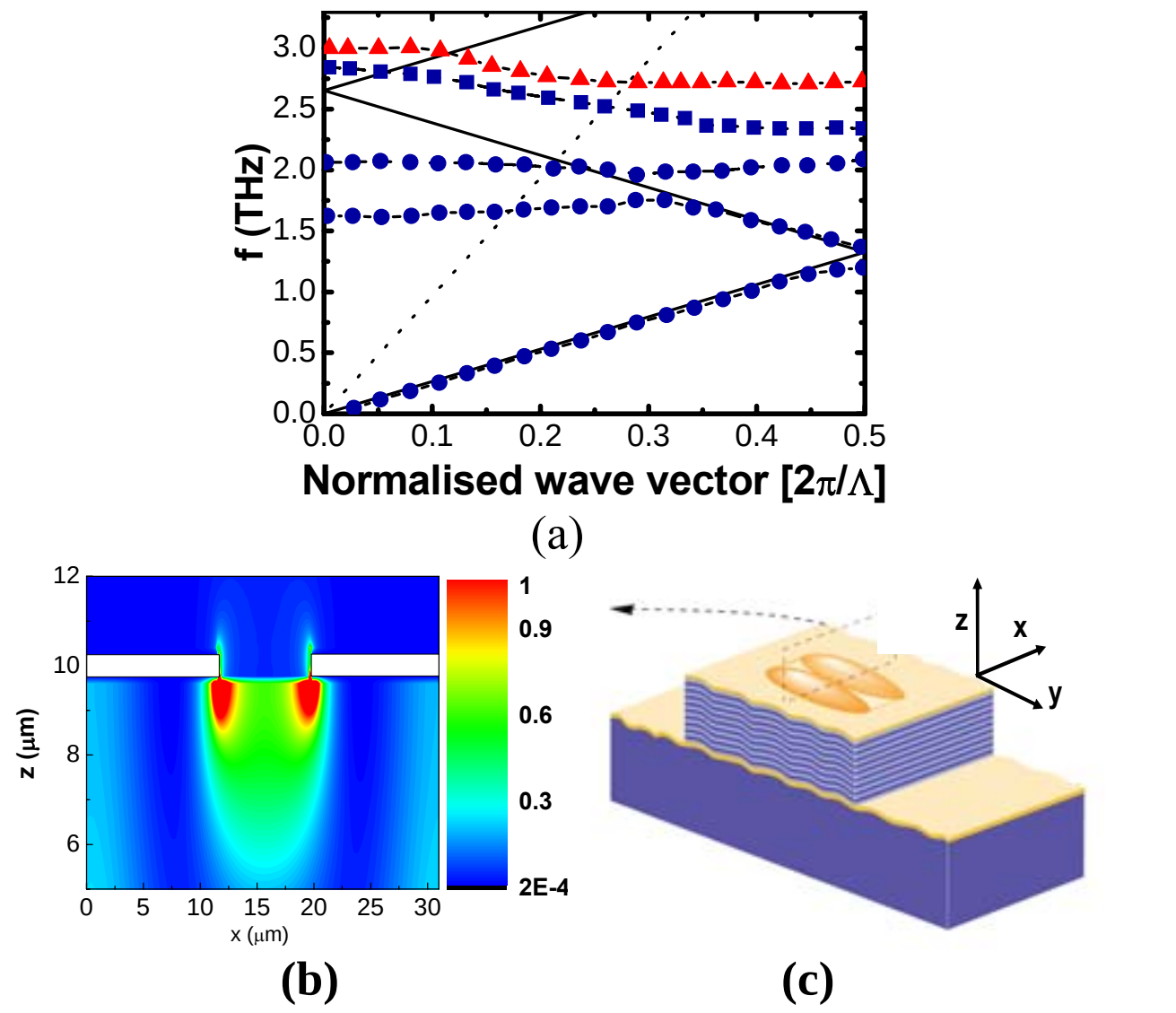

Fig. 4 (a) THz photon modes supported by the device of Fig.1, which is modelled as an superlattice array of slots on the top of an infinite (in x) ridge waveguide. Modes are enumerated in terms of the superlattice wavevector, $2\pi/\Lambda$, where $\Lambda$ = 31 μm is the slot spacing. Squares, the "radiating" mode designed to out-couple radiation when electrically driven as a QCL Red triangles:- the "localised" family of modes. Dotted (solid) lines are light lines in the vacuum (semiconductor). (b) energy distribution, across a single slot in the periodic structure, of the "localised" modes. (c) Schematic 3D distribution of the fields in the "localised" mode concentrated at the edges of the slots in the structure.

almost the same energy (12.5 meV/ ~3 THz) there is another family of "localised" THz modes [fig. 4(a) triangles] whose field distributions are tightly localised at the slot edges, similar to the ultra-confined modes recently reported [17,18] in other sub-wavelength structures. These originate from a vertical ¼ wave "organ–pipe" resonance, with a node at the lower gold layer and an antinode at the slot opening, giving a resonance roughly corresponding to a free-space wavelength $\lambda \sim 4nh$, where n is the semiconductor refractive index and $h \sim 10$ μm the slab thickness. The field localisation means that photon modes on adjacent slots oscillate almost independently, so they are practically mono-energetic and dispersionless in the photonic superlattice plot (fig . 4(a)), and all the THz modes in the family can strongly couple to the electronic ISBT's at the same time. This contrasts with the anti-crossing behaviour previously seen in dispersive 2D systems [11-15]. The field localisation around the slot edge also gives very weak outcoupling to the free-space THz modes, raising the Q factor to ~1100, compared with ~57 for the "radiating" modes.

The computed volume of the "localised" mode, V~ 496 $(\mu m)^3$ , is only $\sim\lambda^3/2000$ of the $\lambda \sim 100$ μm free space wavelength and ~1/50 of $\lambda^3$ in the semiconductor material. It's frequency resonates closely with the modelled $E_{12} \sim 15$ meV (fig. 1) ISBT energy and its electric field is mainly vertically polarised, so it couples strongly to the vertically polarised $z_{12}$ of the ISBT. Also, it's half-height energy density (fig. 4b) is only ~0.96 μm below the semiconductor-air interface, so overlaps well with the ~1μm penetration depth of the $\omega_{NIR}$ incoherent pump light from the the Ti:sapph laser[16].

We compute $\hbar\Omega_{VR}$ for the "localized" photon mode by equating the classical stored electromagnetic energy, $V\varepsilon_r\varepsilon_0 E^2_{vac}$, with the quantum photon ground state energy, $\hbar\omega_{THz}/2$, to give a mean zero-point vacuum field of E ~ 142 V $m^{-1}$. The interaction of a single electron ISBT oscillator with this field will give $\hbar\Omega_{VR} = 2Ee\, z_{12}$, which, with $\varepsilon_r = 13.3$ [16] , $z_{12} \sim 2.3$ nm, and $\hbar\omega_{THz}$ =15 meV gives 6.3 x $10^{-7}$ eV. A factor f=0.92 of the mode energy lies inside the semiconductor, and N~2.4 x $10^6$ ISBT electrons lies within this volume, giving a total coupling energy of $\hbar\Omega_{VR} = 2Ee\, z_{12}\, N^{1/2} \sim 1.0$ meV. Even with no photoexcitation this is some ~7% of the ISBT energy, and will increase further under the experimental conditions. Assuming an interband carrier recombination time of ~1 nsec, ~1mW of absorbed laser power would triple the local electron concentration and increase $\hbar\Omega_{VR}$ by $\sim\sqrt{3}$.

This $\hbar\Omega_{VR}$ value exceeds both the <0.5meV upper bound to the linewidth of the excitation responsible for the SB generation and the ~15 μeV modelled linewidth of the localised photon mode. This confirms the SC nature of the electron-photon system, i.e. the SB's arise from coherent scattering mechanism from polaritons whose linewidths lie between [6] the ~meV ISBT linewidth and the ~15 μeV localised photon mode linewidth.

At higher optical excitation levels (not shown), the $\lambda \sim 100$μm device n=1 SB conversion efficiency peaks at ~ 5 x $10^{-5}$ [at ~ 1 mW input power], and then drops, because of sample heating (fig. 2(c)].

There are strong parallels between this SC system, and previous atom-cavity studies [19] , where coherent output radiation was seen without the system needing to be driven into population inversion. Unfortunately, our attempts to detect the emitted optical component of the THz polaritons produced in these experiments (i.e. to demonstrate a THz analogue of a so called "inversionless laser") were frustrated by the poor sensitivity of current THz detectors, and the very weak coupling of the "localised" modes to the outside world.

That said, we believe that this effect may prove more useful as a simple, passive coherent optical mixing devices than as a source of THz radiation. Although the conversion efficiencies and operating temperatures are low at the moment, this effect still has potential for frequency-shifting e.g. an optical bit stream by a fixed frequency interval that can be tightly specified at the design stage and would be data-transparent and operate across the full optical telecommunications bandwidth. Moving to the [In,Al,Ga],As materials system and optimising $\hbar\Omega_{VR}$ by judicious choice of doping levels, THz mode shapes and ISBT $z_{12}$ values may move the operating wavelengths, temperatures and efficiencies towards technologically useful values.

Helpful conversations with Paul Eastham are gratefully acknowledged. This work was supported by the Engineering and Physical Sciences Research Council (EPSRC), and by the US Air Force Office of Scientific Research AFOSR.

---

[1] M S Skolnick, T A Fisher and D M Whittaker, Semicon. Sci . Tech, **13**, 645 (1998).
[2] B S Williams, S Kumar, Q Hu and J L Reno, Opt. Express **13**, (9), 3331 (2005).
[3] B S Williams, *et al.* , Appl. Phys. Lett., **88**, 261101 (2006).
[4] S Kumar, *et al.* Opt. Express, 15 , 113. (2006).
[5] SA Maier, Opt. Express **14**, 1957 (2006).
[6] Electroluminescence (EL) spectra from test diodes had ~4meV linewidths, but these came from "diagonal" ISBT's with strong interface broadening arising from the high electric fields needs for the measurements. ~2 meV ISBT linewidths were seen in reflectance measurements of Planar samples with similar (3.7THz) ISBT energies to the device of fig. 1. [12].
[7] S. S. Dhillon, et.al., Appl. Phys. Lett., **87**, 071101 (2005).
[8] C. Zervos et.al. Appl. Phys. Lett. **89**, 183507 (2006)
[9] D. E. Nikonov , A. Imamoglu, L. V. Butov, and H. Schmidt,. Phys. Rev. Lett. **79**, 4633 (1997).
[10] D. Dini, Phys.Rev.Lett. **90**, 116401 (2003).
[11] E. Dupont, H. C. Liu, S. Schmidt and A. Seilmeier, Appl.Phys.Lett. **79** 4295 (2001).
[12] Y Todorov et al. Phys Rev Lett., **102** 186404 (2009).
[13] A A Anappara, *et al.* Appl. Phys.Lett. **91**, 231118, (2007).
[14] Jonathan Plumridge, Edmund Clarke, Ray Murray and Chris Phillips, Solid State Comm., **146**, 406, (2008).
[15] C Ciuti, G Bastard, I Carusotto Phys. Rev .B **72**, 115303 (2005).
[16] From the point of view of interband absorption and dielectric response, the heterostructure slab region closely approximated n~ 5 x $10^{15}$ $cm^{-3}$ bulk GaAs, with absorption properties as described in Blakemore, JS, J. Appl. Phys. **53**, R123 (1982) .
[17] M A Seo et al. Nat. Photon., **3**, 152 (2009).
[18] Rupert F. Oulton et.al. Nature, **461,** 629 (2009).
[19] J McKeever et al. Nature, **425**, 268. (2003).